\documentclass[twocolumn,showpacs,amsmath,amssymb,prb,superscriptaddress]{revtex4-1}
\usepackage{graphicx}
\usepackage{color}
\begin{document}

\title{Surface Acoustic Waves Probe of the Spin  Phase Transition at $\nu$=2/3 in n-GaAs/AlGaAs structure}

\author{I.~L.~Drichko}
\author{I.~Yu.~Smirnov}
\affiliation{A.~F.~Ioffe Physico-Technical Institute of Russian
Academy of Sciences, 194021 St. Petersburg, Russia}
\author{A.~V.~Suslov}
\affiliation{National High Magnetic Field Laboratory, Tallahassee,
FL 32310, USA}
\author{L.~N.~Pfeiffer}
\author{K.~W.~West}
\affiliation{Department of Electrical Engineering, Princeton
University, Princeton, NJ 08544, USA}

\begin{abstract}
High frequency (ac) conductivity in the single quantum well
AlGaAs/GaAs/AlGaAs with high mobility was investigated by
contactless acoustic methods in the fractional quantum Hall effect
regime in perpendicular and tilted magnetic fields. We studied the
dependence of ac conductivity $\sigma^{ac}=\sigma_1 - i\sigma_2$ on
both the temperature and magnetic field tilt angle. Tilting the
magnetic field relative to the sample surface enabled us to change
the position of the conductivity oscillation minimum at $\nu$=2/3.
We measured the temperature dependence of ac conductivity for each
tilt angle and for the 2/3 state we calculated the activation energy
$\Delta  E$ which was derived by constructing the Arrhenius plot ln
$\sigma_1$ against 1/$T$. Analyzing behavior of the activation
energy in total magnetic field for the filling factor 2/3 we
observed a distinct minimum which can be interpreted as the spin
unpolarized-polarized phase transition.\\[0.1in]
\end{abstract}


\maketitle

\section{Introduction}

In high mobility GaAs/AlGaAs structures the fractional quantum Hall
effect (FQHE) is observed in strong magnetic fields at low
temperatures. In this regime at the filling factor $\nu$ =2/3 the
spin unpolarized-polarized transition occurs with change of the
magnetic field or electron density. In accordance with the composite
fermions (CF) picture the $\nu$  =2/3 state becomes the CF state
with the filling factor $\nu^{\textrm{CF}}$=2. The energy spacing
between CFs Landau levels (LLs) is determined by a little portion of
the Coulomb energy $\alpha_c E_c=\alpha_c e^2/\varepsilon  l_B
\propto \sqrt{B}$, where $l_B = \sqrt{\hbar/eB}$ is the magnetic
length, $\alpha_c \ll 1$ is the critical parameter [1]. Each CF LL
could in turn be spin-split into two levels separated by the Zeeman
energy $E_Z= g^{\textrm{CF}} \mu_B B \propto B$, where  $\mu_B$ is
the Bohr magneton, and $g^{\textrm{CF}}=g=$-0.44. The energy gap
between nearest CF LLs with different spin orientation at
$\nu^{\textrm{CF}}$=2 can be expressed as $\Delta E=\alpha_c E_c-g
\mu_B B$. Because of different scaling of the energies $E_Z$ and
$E_C$ with $B$, CF LLs could cross at some critical magnetic field
$B_c$, resulting in $\Delta E \rightarrow 0$. Thus, the two CF LLs
with the same spin orientation could turn to be under the Fermi
level, i.e. the system becomes spin-polarized.

This spin phenomenon has been investigated in numerous papers by
different methods [1-16]. In Fig.1 we have summarized the available
data on the spin transition critical magnetic field $T_¬$. It is
seen from the figure that $T_¬$ increases with the electron density
in the sample [2-5,13,14].

\begin{figure}[h!]
\centering
\includegraphics[height=5.5cm]{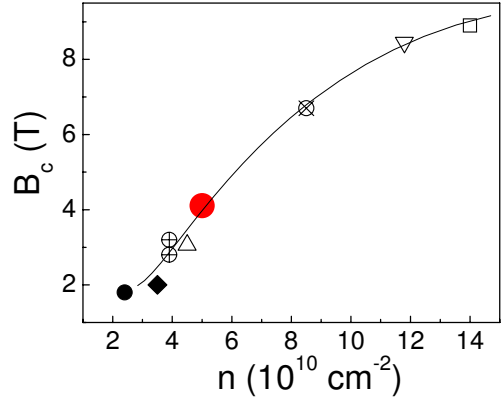}
%\hfill
%\includegraphics[height=5cm]{fig1b.eps}
\caption{The transition critical field vs the electron
density following the results of Ref.2
($\oplus$), Ref.3 ($\bullet$), Ref.4 ($\bigtriangledown$), Ref.5 ($\bigtriangleup$),
Ref.13 ($\otimes, \square$), Ref.14 ($\blacklozenge$),
and the current paper (\textcolor[rgb]{0.98,0.00,0.00}{$\bullet$}).The line is a guide for the eye.
 \label{fig1_Buenos}}
\end{figure}

A study of the FQHE using acoustic methods at $\nu$=2/3 was
implemented in Ref.[4], where the acoustic velocity shift was
measured, but no transition effects were observed. Earlier we also
used the acoustic methods [12] for study the ac conductivity of the
fully spin-polarized state of composite fermions at $\nu$=2/3 in the
GaAs/AlGaAs with electron density $n \approx$ 2 $\times 10^{11}$
 cm$^{-2}$ and mobility $\mu \approx$ 1.5 $\times 10^6 $ cm$^2$/Vs.
Rather high concentration of our sample did not allowed us to
explore this transition directly, since the $\nu$ =2/3 state in our
experiment was at $B$=12 T, while the spin transition occurs at
lower magnetic fields. That is why we have determined just the
dependence of the energy gap $\Delta E$ on magnetic field in the
fully spin polarized state, where it turns out that $\Delta E
\propto B$.

In this paper, we report the activation energy study of the 2/3
state using the Surface Acoustic Wave (SAW) contactless technique in
a high mobility heterostructure GaAlAs/GaAs in tilted magnetic
fields. The concentration of electrons $n \approx$ 5.5 $\times
10^{10}$
 cm$^{-2}$
provides position of the oscillation corresponding to $\nu$=2/3 at
$B_{\perp} \approx$3.2 T, which is rather low. Accordingly with
Fig.1, we shall be possibly able to determine the energy gap in the
transition region by tilting the magnetic field. Thus, for the first
time we would apply the acoustic methods to observe the spin
unpolarized-polarized phase transition at $\nu$=2/3.

\section{Experiment}

In the experiments we used the "hybrid" configuration of the
contactless acoustic method: the surface acoustic wave, excited by
interdigital transducers, propagates along the surface of the
piezodielectric lithium niobate, and the studied structure is
pressed onto the surface of the LiNbO$_3$ (see Fig.2a). The electric
field produced by the SAW interacts with the carriers in 2D channel.
Thus, the attenuation  $\Gamma$  and velocity $v / v_0$ of the
acoustic wave are affected by the conductivity of the 2DEG. In this
"hybrid" setup no deformation is transmitted into the sample. The
technique was first employed for GaAs/AlGaAs structures in [17]. A
detailed acoustic study of the FQHE has been performed in [18].
\begin{figure}[h!]
\centering
\includegraphics[height=5cm]{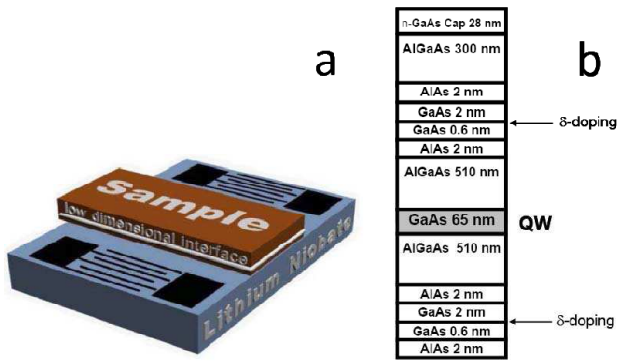}
%\hfill
%\includegraphics[height=5cm]{fig1b.eps}
\caption{A sketch of the acoustic experimental setup (a)
and a cross-section of the studied sample (b).
 \label{fig2}}
\end{figure}

The samples grown by molecular beam epitaxy has a 65 nm wide
symmetrically doped GaAs/AlGaAs quantum well (QW) with the density
of $n \approx$ 5.5 $\times 10^{10}$
 cm$^{-2}$ and mobility of  $\mu =$8.5$ \times 10^6
$cm$^2$/Vs and contained undoped Al$_{0.24}$Ga$_{0.76}$As spacer
layers and Si $\delta$-doped layers (see Fig.2b). The low
temperature measurements in the perpendicular field were done at a
dilution refrigerator while tilting the sample was carried out in a
$^3$He system equipped with a one-axis rotator.

\section{Experimental results and discussion}

The measurements of the SAW attenuation  $\Gamma$  and velocity
change $\Delta v/v_0$ in this system were done at the frequency of
86 MHz in magnetic fields B up to 18 T and in the temperature range
of 0.1 - 1.6 K. Figure 3 shows the dependences of the SAW
attenuation change $\Gamma$  (top panel) and the change of the SAW
velocity $\Delta v/v0$ (middle panel) on the magnetic field $B$
applied along the QW normal z for different temperatures in the QHE
regime. One can see that both the absorption coefficient and the
velocity change exhibit IQHE and FQHE type oscillations in the
magnetic field.
\begin{figure}[h!]
\centering
\includegraphics[height=10cm]{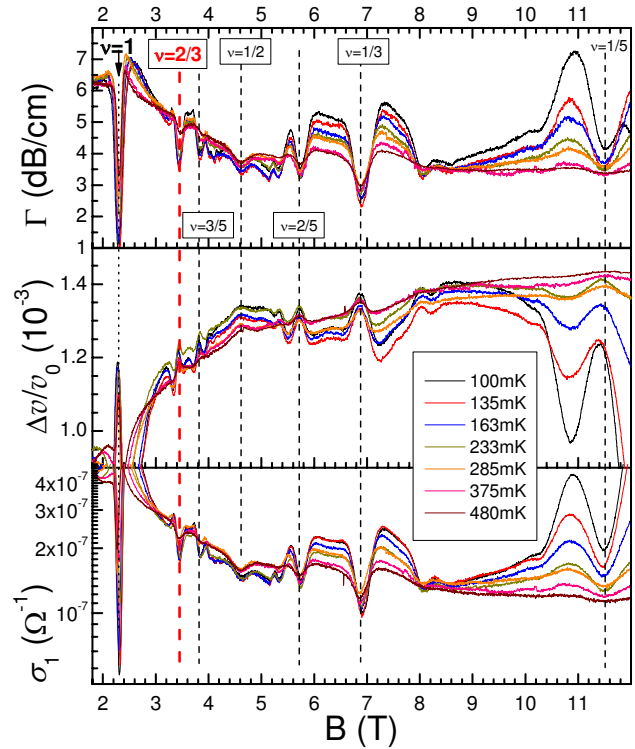}
%\hfill
%\includegraphics[height=5cm]{fig1b.eps}
\caption{Temperature evolution of the SAW attenuation and velocity shift as well as the real
 part of the ac conductivity $\sigma_1$ in the QHE regime in the
perpendicular magnetic field, $f$=86 MHz. The values of the filling factors $\nu$  are shown by vertical dash lines.
 \label{fig3}}
\end{figure}

From the experimentally measured values of the SAW absorption Gamma
and the relative change of the SAW velocity $\Delta v/v_0$, one can
calculate the real $\sigma_1$ and imaginary $\sigma_2$ components of
the high-frequency conductivity $\sigma^{ac}=\sigma_1 - i\sigma_2$
in the electron channel using the Eqs.(1) and (2) of Ref.12. Below,
we focus only on the real part of the ac conductivity since analysis
of this conductivity enables us to determine the parameters of the
energy spectrum. The corresponding dependence of $\sigma_1$ on the
magnetic field for different temperatures at $f$=86 MHz is presented
in the bottom panel of Fig.3. The magnetic field dependence of the
ac conductivity also contains a rich oscillations pattern. At the
fields higher than 3 T $\sigma_1 (B)$ manifests pronounced FQHE
oscillations, which are caused by the formation of CF Landau levels
in the two-dimensional electron gas and are similar to the
oscillations of the magnetoconductivity in the quantum Hall effect
observed in dc transport measurements.

In this paper we aimed at the activation gap studies. To conduct
such measurements properly one indeed needs to find at first the
temperature domain where the conductivity is activated. Compassed in
this way we studied the behavior of ac magnetoconductivity as a
function of temperature down to 100 mK. We found that the ac
conductivity at $\nu$=2/3 is reliably activated at $T>$0.3 K.

We also carried out measurements of the acoustoelectric effects and
calculated $\sigma_1$ in a tilted magnetic field, being focused
mainly on the 2/3 state which corresponds to the normal component of
total magnetic field $B_{\perp} =B_{\textrm{TOTAL}} \cos(\Theta)
\approx$3.2 T. We measured the angle by tracing position of the most
pronounced oscillations, namely the one of $\nu$=1. Tilting the
magnetic field relative to the sample surface enabled us to change
the position of the conductivity oscillation minimum at $\nu$=2/3.

We measured the temperature dependence of  $\Gamma$  and $\Delta
v/v_0$ in the range of 0.3 to 1 K for each tilt angle and derived
$\sigma_1$. The activation energy $\Delta E_{2/3}$ was derived by
constructing the Arrhenius plot $\ln \sigma_1$ against 1/$T$
assuming that the conductivity $\sigma_1 \propto \exp(-\Delta
E_{2/3}/2k_B T)$. Dependence of the activation energy for $\nu$=2/3
$\Delta E_{2/3}$ on total magnetic field is illustrated in Fig.4. As
one can see, the energy gap in the magnetic field for filling factor
2/3 passes through a distinct minimum, which can be interpreted as
the unpolarized-polarized phase transition when a crossing of the
composite fermion's Landau levels with different spin directions at
$\nu^{\textrm{CF}}$=2 occurs. At the crossing between two CF LLs the
energy gap should disappear at the transition point. However, as
seen in Fig.4 $\Delta E_{2/3}$ does not approach the zero. This
anticrossing could evolve as a result of the electron exchange
interaction as discussed in Ref.[4].
\begin{figure}[h!]
\centering
\includegraphics[height=6cm]{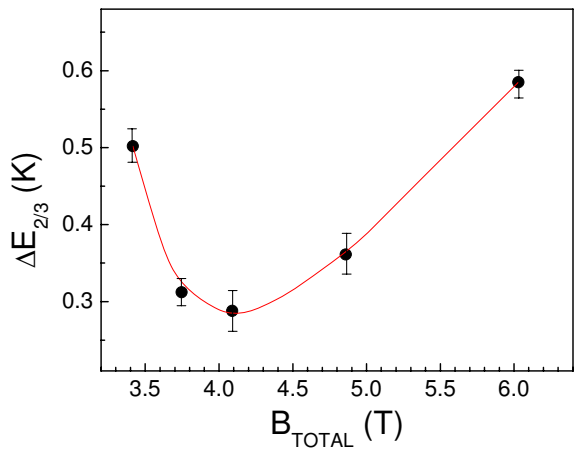}
%\hfill
%\includegraphics[height=5cm]{fig1b.eps}
\caption{Dependence of the activation gap for $\nu$=2/3 on total magnetic field. The line is a guide to the eye.
 \label{fig4}}
\end{figure}

It is known that if a quantum well is wide the Coulomb interaction
could be weakened. As a result, at the layer thickness exceeding
3$l_B$  the FQHE quickly collapses (see [19] and refs therein). In
our sample the QW width is 65 nm, which several times exceeds the
magnetic length at $B$=4T. However, using the acoustic technique we
were able not only to observe a rich FQHE pattern, but also to
measure the temperature dependence of conductivity at $\nu$=2/3, as
well as to determine the dependence of the energy gap for
$\nu^{\textrm{CF}}$=2 on the magnetic field. These measurements
allowed us to observe the transition from unpolarized to polarized
FQHE states at the CF filling factor $\nu^{\textrm{CF}}$=2. Critical
field of this transition $B$=4.2 T is indicated in Figure 1 by the
red dot. If we assume that CF g-factor  $|g^{\textrm{CF}}| = |g|
=$0.44, then in the spin transition point at $B_c$ we can estimate
the value of the critical parameter $\alpha_c =E_z/E_c=$0.012, which
is close to the value of $\alpha_c \approx$0.008, determined in the
Ref.[1] for the 65 nm wide quantum well for $\nu$=2/3.

In conclusion, we performed Surface Acoustic Waves contactless
measurements of the activation energy of the 2/3 state in tilted
magnetic fields, thus, having found the critical field of the spin
transition in dilute GaAs/AlGaAs structure with wide quantum well.

\section*{Acknowledgments}

The authors would like to thank E. Palm, T. Murphy, J.-H. Park, and
G. Jones for technical assistance. NHMFL is supported by NSF
Cooperative Agreement No. DMR-1157490, the State of Florida, and the
U.S. Department of Energy. The work at Princeton was partially
funded by the Gordon and Betty Moore Foundation through Grant
GBMF2719, and by the National Science Foundation MRSEC-DMR-0819860
at the Princeton Center for Complex Materials.

\end{document}